\documentclass{nature}
\usepackage{bm}
\usepackage[colorlinks=true, letterpaper=true, pdfstartview=FitV, linkcolor=blue, citecolor=blue, urlcolor=blue]{hyperref}
\usepackage{siunitx}
\usepackage{amssymb}
\usepackage{xcolor}
\usepackage{mathtools}
\DeclarePairedDelimiter\ceil{\lceil}{\rceil}

\title{Probing Landau levels of strongly interacting massive Dirac electrons in layer-polarized MoS$_2$}
\author{Jiangxiazi Lin$^1$, Tianyi Han$^1$, Benjamin A. Piot$^2$, Zefei Wu$^1$, Shuigang Xu$^1$,
Gen Long$^1$, Liheng An$^1$, Patrick Ka Man Cheung$^4$, Peng-Peng Zheng$^4$,
Paulina Plochocka$^3$, Duncan K. Maude$^3$, Fan Zhang$^4$*, \& Ning Wang$^1$*}
\begin{document}
\maketitle
\begin{affiliations}
 \item Department of Physics and Center for Quantum Materials,
 The Hong Kong University of Science and Technology, Clear Water Bay, Hong Kong, China
 \item Laboratoire National des Champs Magn\'{e}tiques Intenses,
 LNCMI-CNRS-UGA-UPS-INSA-EMFL, F-38042 Grenoble, France
 \item Laboratoire National des Champs Magn\'{e}tiques Intenses,
 LNCMI-CNRS-UGA-UPS-INSA-EMFL, F-31400 Toulouse, France
 \item Department of Physics, The University of Texas at Dallas, Richardson, Texas 75080, USA\\
 $^*$E-mails: zhang@utdallas.edu; phwang@ust.hk
\end{affiliations}

\begin{abstract}
Monolayer transition metal dichalcogenides are recently emerged 2D electronic systems
with various novel properties, such as spin-valley locking, circular dichroism, valley Hall effects, Ising superconductivity~\cite{wang2012electronics,xu2014spin,Mak2016NP,li2013unconventional,lu2015evidence}.
The reduced dimensionality and large effective masses further produce unconventional many-body interaction effects.
Although recent hole transport measurements in WSe$_2$ indicate strong interactions in the valence bands~\cite{xu2017odd,movva2017density,gustafsson2017ambipolar,xu2017probing},
many-body interaction effects, particularly in the conduction bands, remain elusive to date.
Here, for the first time, we perform transport measurements up to a magnetic field of $29$~T
to study the massive Dirac electron Landau levels (LL) in layer-polarized MoS$_2$ samples
with mobilities of $22000$~cm$^2$/(V$\cdot$s) at $1.5$~K and densities of $\sim10^{12}$~cm$^{-2}$.
With decreasing the density, we observe LL crossing induced valley ferrimagnet-to-ferromagnet transitions,
as a result of the interaction enhancement of the g-factor~\cite{zhu2003spin,spivak2010colloquium} from $5.64$ to $21.82$.
Near integer ratios of Zeeman-to-cyclotron energies, we discover LL anticrossings due to the formation of
quantum Hall Ising ferromagnets~\cite{nicholas1988exchange,jungwirth1998magnetic,girvin2000spin},
the valley polarizations of which appear to be reversible by tuning the density or an in-plane magnetic field.
Our results provide compelling evidence for many-body interaction effects
in the conduction bands of monolayer MoS$_2$ and establish a fertile ground for
exploring strongly correlated phenomena of massive Dirac electrons.
\end{abstract}

\begin{figure}
\includegraphics[width=1\columnwidth]{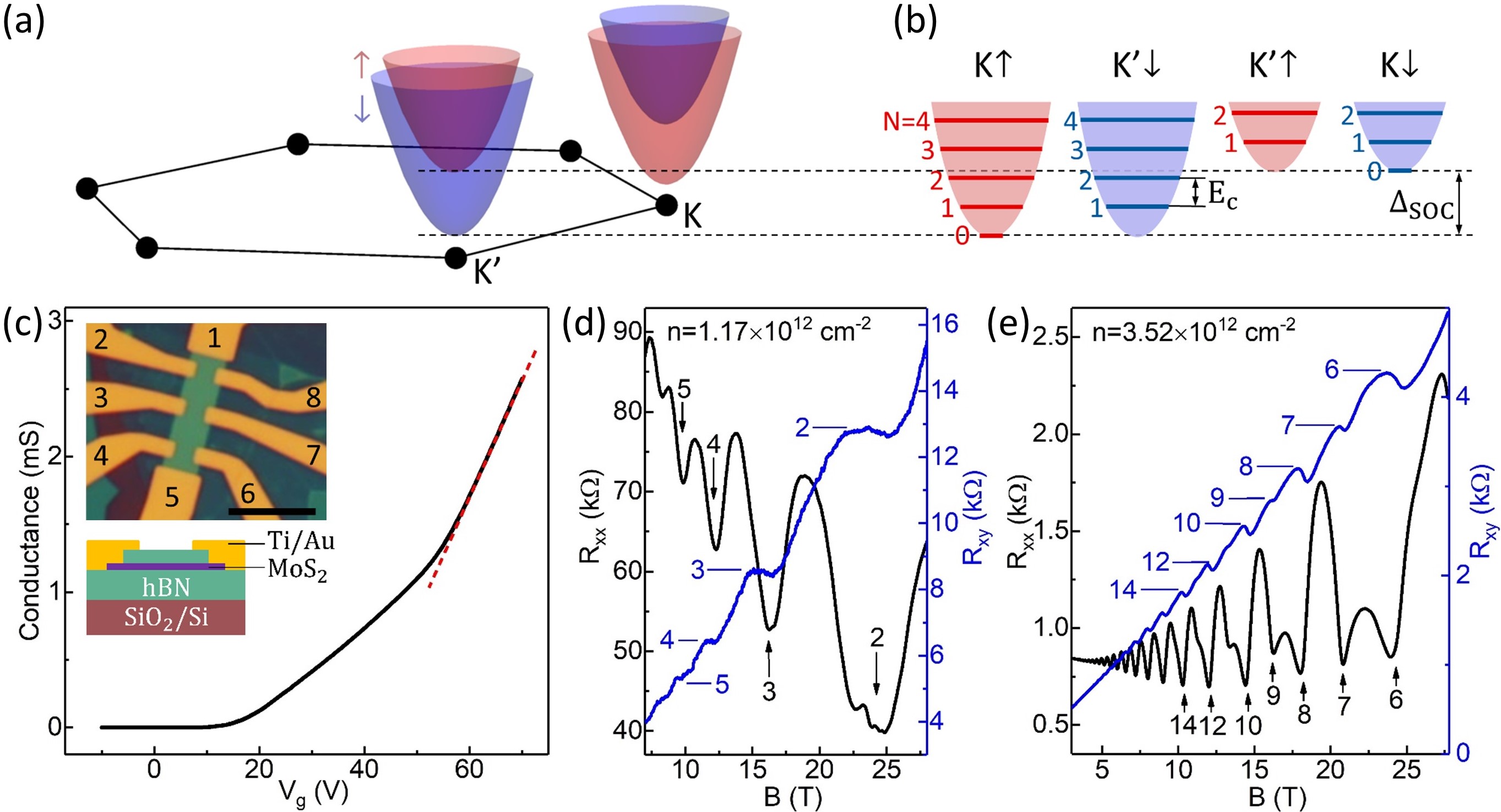}
\caption{{\bf Band structure, device structure, field-effect conductance,
and quantum oscillations of the layer-polarized MoS$_2$.}
(a) Hexagonal Brillouin zone of monolayer MoS$_2$ and its conduction band minima at K/K' points, with a splitting between the spin up (red) and spin down (blue) subbands.
(b) Spin-valley locked conduction-band Landau level structure of massive Dirac electron bands,
with Zeeman effect not considered.
$\Delta_{SOC}$ is the spin splitting in each valley at zero magnetic field.
(c) Four-probe FET conductance at 1.5~K. The red dashed line indicates an FE mobility
of 22000~cm$^2/(\mbox{V}\cdot\mbox{s})$.
Inset: the optical image and cross-section of the device, with a scale bar of 8~\si{\micro\meter}.
(d) Magneto (black) and Hall (blue) resistance measured at 1.2~K at a low density,
featuring quantized plateaus in Hall resistance.
(e) The same as (d) but at a higher density, displaying a 2-to-1 valley degeneracy lifting.
\label{fig1}}
\end{figure}

Semiconducting transition metal dichalcogenides (TMDCs) such as MoS$_2$, MoSe$_2$, WS$_2$, and WSe$_2$
form a family of layered 2D materials exhibiting novel electronic and optical properties~\cite{wang2012electronics,
xu2014spin,Mak2016NP,mak2014valley,lu2015evidence,wu2014piezoelectricity,zhang2014electrically}.
In monolayer TMDCs, the inversion symmetry is strongly broken, leading to a massive Dirac band structure
at the K/K' valleys with nontrivial Berry phase effects~\cite{xu2014spin,Mak2016NP,li2013unconventional}.
The spin-orbit coupling (SOC) further splits the two spin subbands in each valley,
whereas the spins are quantized along the out-of-plane direction due to the mirror-plane symmetry.
Thus, below the energy of intra-valley spin splitting,
the subbands are valley degenerate and spin-valley locked (Fig.~\ref{fig1}a).
Such a remarkable band structure gives rise to the intriguing circular dichroism~\cite{wang2012electronics,xu2014spin,Mak2016NP},
valley Hall effects~\cite{mak2014valley}, Ising superconductivity~\cite{lu2015evidence},
and anomalous Landau level (LL) structure~\cite{li2013unconventional,Mak2016NN} (Fig.~\ref{fig1}b).

By contrast, intrinsic bilayer MoS$_2$ has a restored inversion symmetry and
enjoys a spin degeneracy of opposite layers~\cite{wu2012electrical,lee2015electrical}.
Because of the vanishing interlayer coupling at the conduction band edge at the K/K'
points~\cite{liu2013three}, a backgate voltage $V_g$ can easily induce a potential difference between
the two layers~\cite{wu2012electrical,lee2015electrical,supp}, thereby
breaking the inversion symmetry and polarizing the electrons to one of the two layers
when Fermi level is lower than the potential difference.
Therefore, the low energy electron band structure for such a layer-polarized MoS$_2$ (Fig.~\ref{fig1}a)
resembles that of monolayer MoS$_2$~\cite{xu2014spin,Mak2016NP,li2013unconventional},
for which it is technically more difficult to achieve high-mobility samples.

When subjected to a perpendicular magnetic field $B$, the massive Dirac electrons
are quantized into anomalous LLs (Fig.~\ref{fig1}b) with a cyclotron gap $E_{c}=\hbar eB/m^*$,
where $\hbar$ is the reduced Planck constant, $e$ is the electron charge,
and $m^*$ is the effective mass. Notably, the $N=0$ LLs exist in only one valley.
Below the energy of SOC induced spin splitting $\Delta_{SOC}$,
the $N\neq0$ LLs are valley degenerate and spin-valley locked (Fig.~\ref{fig1}b).
A valley Zeeman effect~\cite{Mak2016NP,wu2016even} can lift the degeneracy
by producing a gap $E_Z=g_{LL}\mu_B B$, where $g_{LL}$ is the effective g-factor in the LL structure,
and $\mu_B$ is the Bohr magneton. The ratio of Zeeman-to-cyclotron energies $E_Z/E_{c}$
can be strongly enhanced by electron-electron interactions,
and the LLs of different valleys would cross each other if $E_Z/E_{c}>1$.
Based on such LL crossings or the more tantalizing anticrossings, here
we examine the strongly interacting behaviors of massive Dirac electrons in the layer-polarized MoS$_2$.

A bilayer MoS$_2$ flake was first exfoliated onto a silicon wafer, then identified by optical contrast,
and subsequently encapsulated by two insulating hexagonal boron nitride (hBN) layers.
The hBN encapsulation protects
the MoS$_2$ from impurity contamination and degradation~\cite{wang2013one,xu2016universal}.
The sandwich structure was patterned
and contacted by Ti/Au electrodes~\cite{wang2013one,xu2016universal}.
Figure~\ref{fig1}c shows the vertical cross section and optical image of the device.
The electrodes are labeled from 1 to 8, among which 1 and 5 were used as current injection,
2 and 4 as longitudinal voltage probes, and 2 and 8 as Hall voltage probes.
The encapsulated device lies on a SiO$_2$/Si wafer, to which the $V_g$ was applied.
All data were obtained by a standard low-frequency lock-in technique at cryogenic temperatures.
We fabricated and measured three samples with the same structure, labelled A, B, and C.
They show consistent behaviors, and only results from sample C are presented here.
(Data of the other two samples are provided in Supplementary Materials~\cite{supp}.)

From the conductance of the {\it n}-type field effect channel at 1.2~K (Fig.~\ref{fig1}c), the mobility is
calculated as $\mu=(\mbox{d}G/\mbox{d}V_g )l/(wC_g )\approx 22000$~cm$^2$/(V$\cdot$s),
where $l=7.6$~\si{\micro\meter} and $w=2.6$~\si{\micro\meter} are
the length and width of the channel, respectively, $C_g$ is the gate capacitance determined
by the linear relation between $V_g$ and the electron density $n$~\cite{supp}.
The high mobility and low contact resistance allow us to observe Shubnikov-de Haas (SdH) oscillations
down to a carrier density as low as $6.3\times10^{11}$~cm$^{-2}$~\cite{supp},
which is hitherto the lowest density for few-layer TMDCs.
At $B>10$~T, the system is driven into the quantum Hall regime (Fig.~\ref{fig1}d):
the onset of plateaus are clearly developed in the Hall resistance ($R_{xy}$),
although the minima remain non-zero in the magneto resistance ($R_{xx}$).

Figure~\ref{fig1}e showcases the measured $R_{xx}$ and $R_{xy}$ at $n=3.52\times10^{12}$~cm$^{-2}$.
The LL filling factors are labeled in blue for $R_{xy}$ by $\nu=(2\pi\hbar/e^2)/R_{xy}$
and in black for $R_{xx}$ by $\nu=2\pi n\hbar/eB$. Evidently, the LLs are doubly degenerate at $B<12.5$~T,
and the degeneracy is lifted at higher $B$. The $2$-to-$1$ degeneracy lifting is consistent with
the expected valley Zeeman effect~\cite{Mak2016NP,wu2016even} of the electrons in the polarized layer.
Moreover, the single frequency $B_F$ obtained from Fourier analysis~\cite{supp} excludes
the possible participation of other subbands. The temperature-dependent SdH oscillations
further yield $m^*=0.55\pm0.08$~$m_0$ without obvious dependence on $n$ or $B$;
this value is comparable to the calculated effective
mass~\cite{wang2012electronics,xu2014spin,Mak2016NP,li2013unconventional}
of monolayer electrons in the K/K' valleys.

\begin{figure}
\includegraphics[width=1\columnwidth]{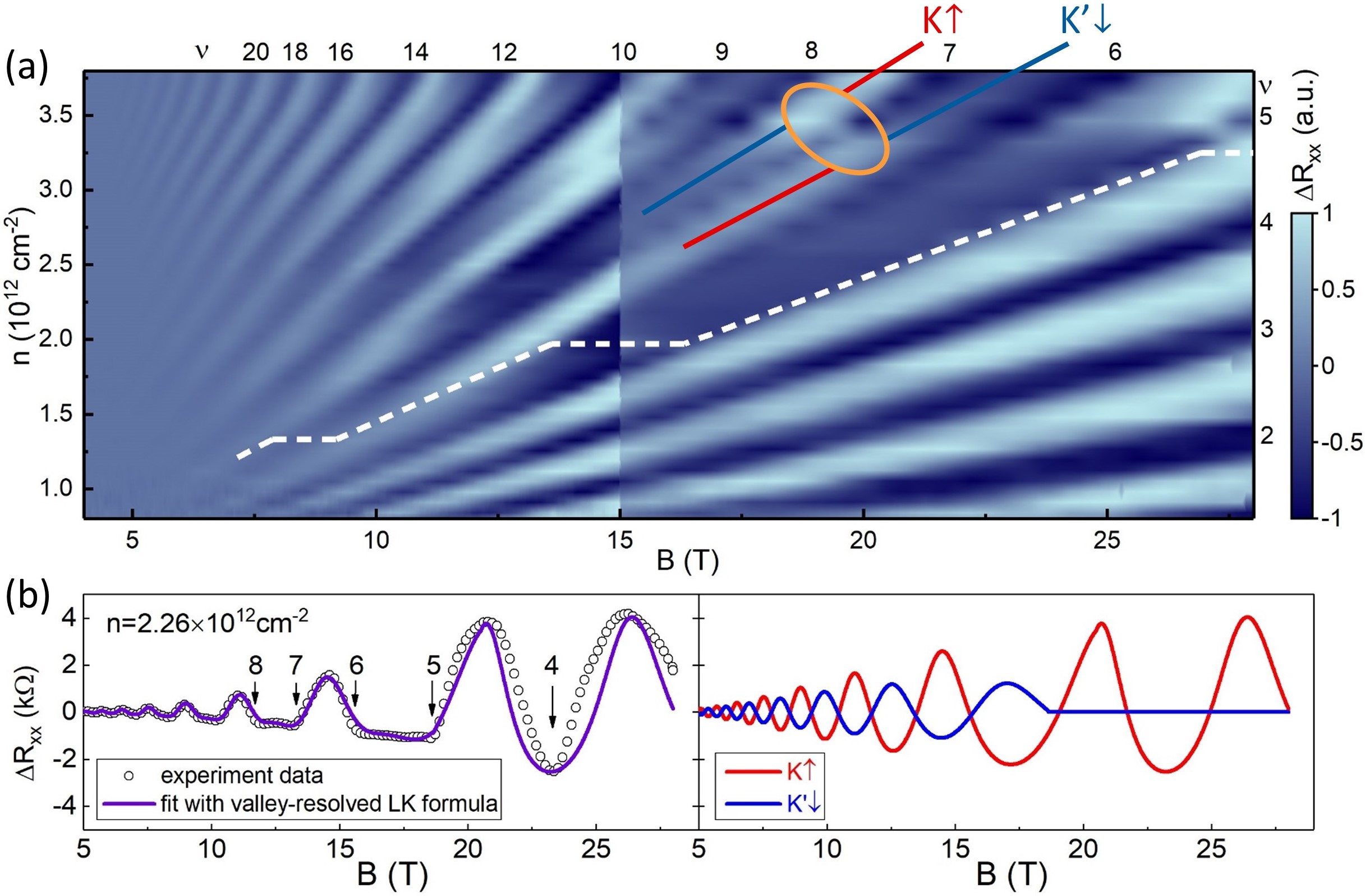}
\caption{{\bf Landau level fan diagram of the layer-polarized MoS$_2$.}
(a) Mapping of quantum oscillation amplitude in the $n-B$ space.
The dark blue regions are $R_{xx}$ minima, related to energy gaps in the density of states,
whose corresponding filling factors are marked along the right and top edges.
The data below and above $15$~T were obtained in a $1.5$~K superconducting cryogenic system
and in a $1.2$~K high magnetic field system, respectively, leading to the minor color discontinuity at $15$~T.
The red and blue solid lines mark a pair of crossing Landau levels with opposite spins and valleys,
and the orange oval is the crossing position.
The white dashed line marks the boundary between valley ferromagnetic and ferrimagnetic regimes.
(b) Left: quantum oscillation data fitted with the valley-resolved LK formula.
Right: two valley components of the fitting.
\label{fig2}}
\end{figure}

To determine the valley susceptibility (related to $E_Z/E_{c}$),
one might suggest to apply an in-plane magnetic field that would
enhance $E_Z$ without affecting $E_{c}$~\cite{xu2017odd,nicholas1988exchange}.
In fact, because of the spin splitting and pinning to the out-of-plane direction,
the LLs are immune to the in-plane Zeeman field.
Thus, the commonly used tilt field method is not applicable to our system~\cite{movva2017density,supp}.
However, for a strongly interacting 2D electron gas, lowering the density $n$ can enhance
$E_Z/E_{c}$~\cite{movva2017density,gustafsson2017ambipolar,xu2017probing,zhu2003spin,spivak2010colloquium}.
It follows that $E_Z /E_{c}$ can be determined by using the LL fan diagram, i.e.,
the mapping of SdH oscillation amplitudes in the $n$-$B$ space.
Figure~\ref{fig2}a plots the LL fan diagram,
in which a white dashed line separates two different regimes.
On the upper left side there are a series of LL crossings and anticrossings with alternating brightness, i.e.,
between the brighter and dimmer levels. For instance, the red and blue lines mark
one pair of anticrossing levels at $\nu=7$, with the anticrossing point highlighted by the orange oval, which will be discussed in the last part of this paper..
By contrast, on the lower right side there are no dimmer levels or crossings.

The alternating brightness (Fig.~\ref{fig2}a) can be attributed to the filling of LLs in different valleys by using
the {\em valley-resolved} Lifshitz-Kosevitch (LK) formula~\cite{pudalov2014probing,maryenko2015spin}:
\begin{equation}
\Delta R_{xx}=2R_0 \sum_{r=1}^{N_r}\sum_{\sigma=K,K'}\frac{r\lambda}{\mbox{sinh}(r\lambda)}\mbox{exp}(\frac{-r\pi\hbar}{E_c\tau_{\sigma}})\mbox{cos}(r\phi_{\sigma})
\end{equation}
where $\lambda=2\pi^2 k_B T/E_c$ is the thermal damping,
and $\tau_{K,K'}$ are the valley-resolved scattering time.
Given the anomalous LL structure for massive Dirac electrons~\cite{li2013unconventional} (Fig.~\ref{fig1}b),
$\phi _{K}=2\pi B_F /B+\pi E_Z /E_{c}$ and $\phi _{K'}=2\pi B_F /B-\pi E_Z /E_{c}-2\pi$.
In Fig.~\ref{fig2}b, the left panel displays the experimental $\Delta R_{xx}$ (open circles)
at $n=2.26\times10^{12}$~cm$^{-2}$ and the fitted $\Delta R_{xx}$ (purple line) using Eq.~(1) with $N_r=10$;
the best fitting yields $E_Z /E_{c}=3.5$. The right panel plots the individual contributions to the fitted $\Delta R_{xx}$
from each valley; while both become stronger with increasing $B$,
those of valley K' electrons are relatively weaker and cease at $\sim18$~T or $\nu=5$.
These findings suggest that the dimmer levels (Fig.~\ref{fig2}a) and the flat $R_{xx}$ region (Fig.~\ref{fig2}b)
correspond to LLs in valley K', and, more significantly, that a valley ferrimagnet-to-ferromagnet
transition occurs at $\sim18$~T or $\nu=5$ for $n=2.26\times10^{12}$~cm$^{-2}$.
In addition, the fitting indicates a valley-dependent scattering time $\tau_K > \tau_{K'}$\cite{supp},
explaining the stronger SdH oscillations in valley K. $\tau_{K, K'}$ obtained by this method have the same order
of magnitude with that obtained by a temperature-dependent analysis~\cite{supp}.

\begin{figure}
\includegraphics[width=1\columnwidth]{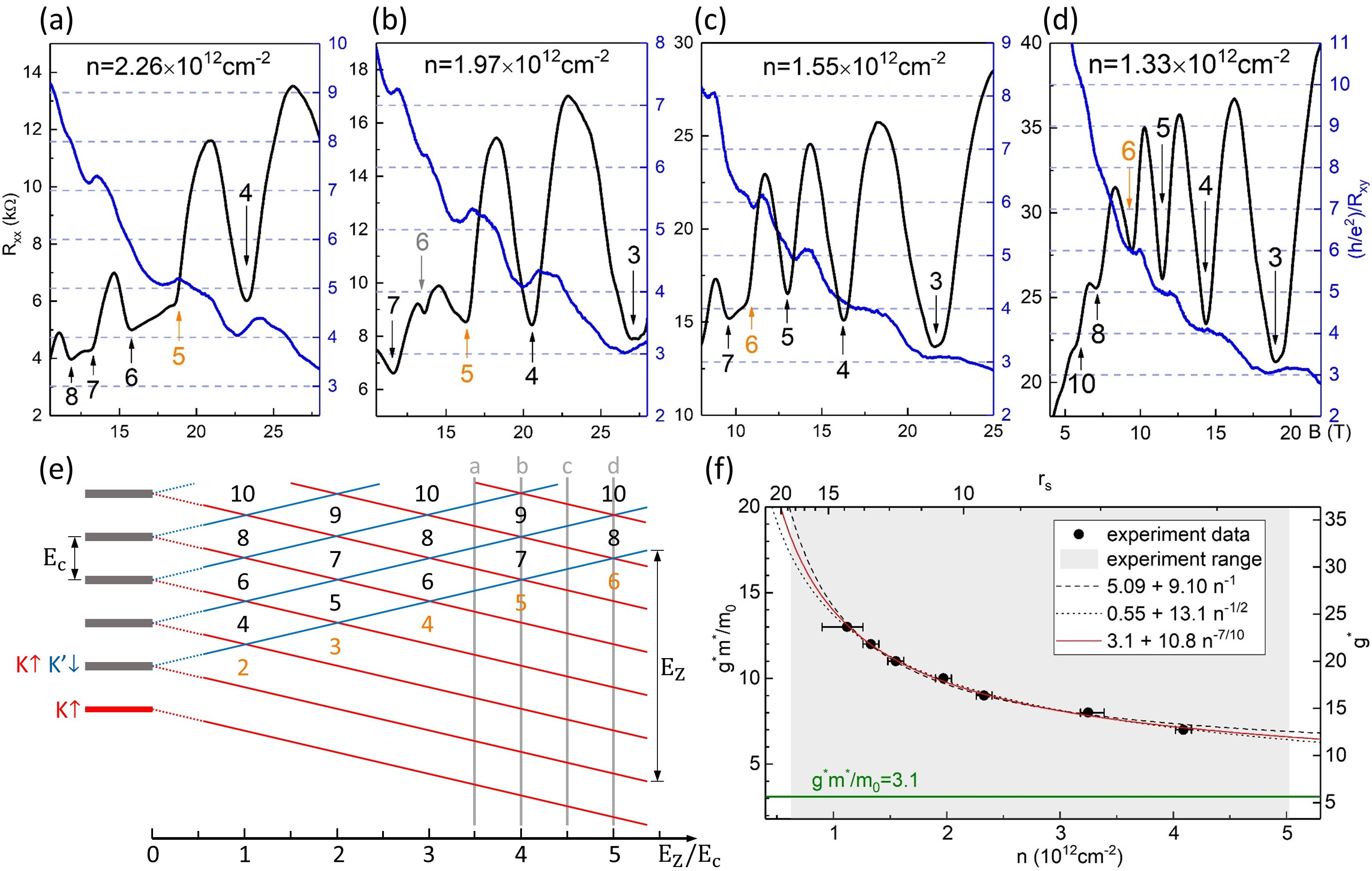}
\caption{{\bf Density dependent valley Zeeman splitting.}
(a)-(d) $R_{xx}$ (black) and $(h/e^2)/R_{xy}$ (blue) at different electron densities.
(e) Schematics of Landau levels with increasing valley Zeeman splitting.
The red and blue lines denote the valley K (spin-up) and valley K' (spin-down) Landau levels.
Four vertical grey lines correspond to the situations in (a) to (d), respectively.
Orange number labels $\nu_c$ that separates the valley ferrimagnetic and ferromagnetic regime.
(f) Experimental values of $g^*$ and $g^*m^*/m_0$ versus $n$ and $r_s$.
The horizontal error bars represent the uncertainty in $V_g$.
Fittings with different formulas are shown.
\label{fig3}}
\end{figure}

When $E_Z /E_{c}$ is a half-integer,
LLs of different valleys are alternating and equally spaced in the valley ferrimagnetic regime,
yielding $R_{xx}$ minima at all integer filling factors. By contrast,
when $E_Z /E_{c}$ is an integer, LLs of different valleys coincide in the valley ferrimagnetic regime,
producing the pronounced signatures only at $\nu= 1, 2, \dots, \nu_c, \nu_c+2, \nu_c+4, \dots$.
At $\nu\leq\nu_c$ the ground state is a valley ferromagnet.
Given the anomalous LL structure for massive Dirac electrons~\cite{li2013unconventional} (Fig.~\ref{fig3}e),
$\nu_c=\ceil[\big]{E_Z /E_c}+1$, where $\ceil[\big]{\;}$ is the ceiling function.
Such principles validate the consistency between the fitted $E_Z /E_c=3.5$ and the observed
ferrimagnet-to-ferromagnet transition at $\nu_c=5$ for $n=2.26\times10^{12}$~cm$^{-2}$ (Fig.~\ref{fig3}a).
Now we use similar transitions in the LL fan diagram to extract the $E_Z /E_c$ values at lower carrier densities.
Since the $R_{xx}$ minima are most striking when $E_Z /E_c$ are integers or half integers,
we identify that $n=$ 1.97, 1.55, and 1.33$\times10^{12}$~cm$^{-2}$
correspond to $E_Z /E_c=$ 4, 4.5, and 5, respectively.
Figures~\ref{fig3}b-\ref{fig3}d display the observed $R_{xx}$ and $R_{xy}$ for these densities,
in good agreement with their characteristic LL structures (Fig.~\ref{fig3}e).

The effective g-factor in the LL structure can be obtained by $g_{LL}=2(E_Z/E_c)/(m^*/m_0)$,
which is $2$ in the large $n$ (non-interacting) limit. In the $B\rightarrow0$ limit,
the effective g-factor for the conduction-band valley Zeeman effect is $g^*=g_{LL}+2m_0/m^*$,
and $2m_0/m^*=3.64$ arises from the valley magnetic moment of massive Dirac bands~\cite{cai2013magnetic}
(which, if $B\neq0$, would lead to the gap between the lowest LLs in different
valleys~\cite{li2013unconventional} (Fig.~\ref{fig1}b)).
The effective valley susceptibility ($\propto g^*m^*$) can be further obtained:
$g^*m^*/m_0=3.1$ ($g^*=5.64$) in the large $n$ limit
and $g^*m^*/m_0=12$ ($g^*=21.82$) at $n=1.33\times10^{12}$~cm$^{-2}$.
(We have assumed $g_{LL}>0$, i.e., the valley with the $N=0$ LL shifts down in energy.
If $g_{LL}<0$, then $g^*m^*/m_0$ would be $0.9$ in the large $n$ limit
and $-12$ at $n=1.33\times10^{12}$~cm$^{-2}$.)
The best fitting in Fig.~\ref{fig3}f gives $g^*m^*/m_0=3.1+10.8n^{-7/10}$,
which may be improved in future experiments that can take into account
the neglected electron-hole asymmetry~\cite{aivazian2015magnetic}
and minority orbitals~\cite{liu2013three}.
The giant valley susceptibility and its enhancement with decreasing $n$
are most likely due to the electron-electron interactions in
our system~\cite{fang1968effects,zhu2003spin,piot2005quantum,spivak2010colloquium}.
The interaction strength can be indicated by the dimensionless Wigner-Seitz radius
$r_s={1}/(\sqrt{\pi n}a^*_B)$~\cite{spivak2010colloquium},
where $a^*_B={4\pi\epsilon\hbar^2}/(m^*e^2)$ is the effective Bohr radius,
and $\epsilon\approx4\epsilon_0$ is the dielectric constant in our sample~\cite{chen2014probing}.
Our studied densities yeild $r_s\sim10$ (Fig.~\ref{fig3}f),
which is indeed a strong interaction regime~\cite{spivak2010colloquium}.

\begin{figure}
\includegraphics[width=1\columnwidth]{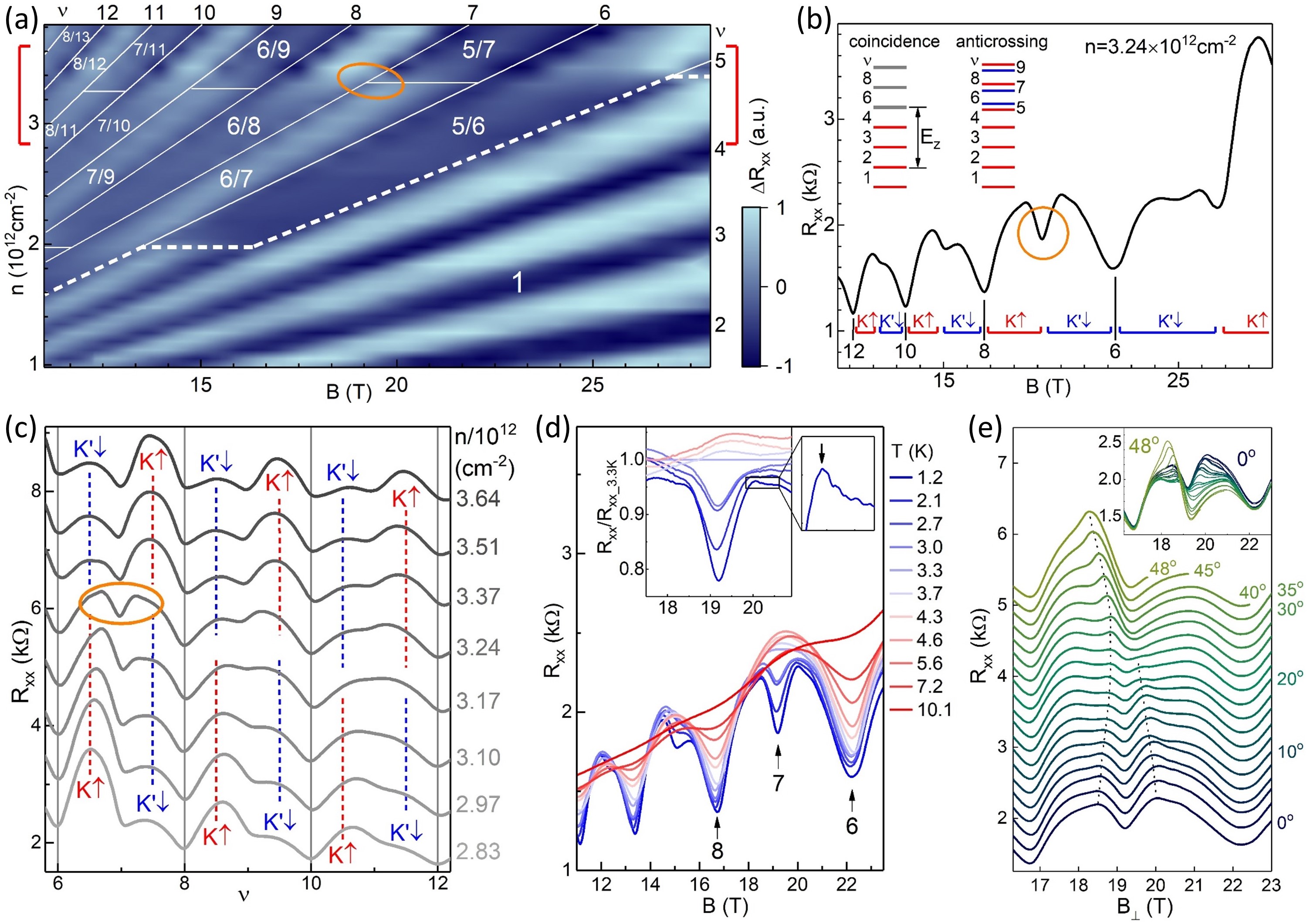}
\caption{{\bf Signatures of Landau level anticrossings.}
(a) Mapping of quantum oscillation amplitude in the $n-B$ space with valley polarizations labeled.
(b) Quantum oscillations showing the valley-resolved Landau levels and their anticrossings.
Inset: the corresponding Landau level structure.
(c) $R_{xx}$ versus $\nu$ at densities in the range marked by the red bracket in (a).
The red and blue dashed lines mark the Landau levels in valley K and K', respectively.
The orange ovals in (a)-(c) mark the same anticrossing at $\nu=7$.
(d) Temperature dependence of the anticrossing signatures in SdH oscillations.
Inset: the normalized $R_{xx}$ near $\nu=7$.
(e) Tilt angle dependence of the anticrossing at $\nu=7$, offset for clarity.
The dotted traces are guides to the eye for the positions of enhanced peaks in $R_{xx}$.
Inset: the non-offset data.
\label{fig4}}
\end{figure}

At those densities with integer $E_Z/E_c$, LL anticrossings are also observed,
e.g., evidenced by the secondary $R_{xx}$ minima at $\nu=$ 5, 7, 9, and 11 in Fig.~\ref{fig4}b
for $n=3.24\times10^{12}$~cm$^{-2}$ ($E_Z/E_{c}=3$).
By applying our established result that the LLs in valley K enjoy larger $R_{xx}$ peaks to Fig.~\ref{fig4}b,
we can directly obtain the depicted LL structure.
This allows us to label the valley polarizations of the quantum Hall states in Fig.~\ref{fig4}a,
as defined by $(n_K-n_{K'})/(n_K+n_{K'})$ with $n_{K,K'}$ is the number of electrons in valley K/K'.
Figure~\ref{fig4}c highlights a strong anticrossing at $\nu=7$ (orange circle):
the two anticrossed LLs of different valleys swap their relative positions abruptly
with varying the density (dotted traces). Such LL anticrossings are most likely due to the formation of
quantum Hall Ising ferromagnets driven by the exchange interaction~\cite{nicholas1988exchange,jungwirth1998magnetic,girvin2000spin}.
This picture is consistent with the observation that the anticrossings at $\nu=$9 and 11 are less pronounced,
as the exchange interaction strength decreases with decreasing $B$ or increasing the LL orbital index.
Note that LL anticrossing is also observed for another density $n=1.97\times10^{12}$~cm$^{-2}$ ($E_Z/E_{c}=4$), as the dip at $\nu=6$ (Fig.~\ref{fig3}b).

Figure~\ref{fig4}d shows the temperature dependence of the anticrossings.
Clearly, the anticrossing gap at $\nu=7$ disappears at $\sim3.3$~K,
whereas the single-particle gap at $\nu=6$ disappears at a much higher temperature $\sim10.1$~K.
This provides further evidence for the many-body origin of the anticrossings.
Moreover, the normalized $R_{xx}$ at $\nu=7$ exhibits a small peak at the shoulder,
as pointed out by the arrow in the inset of Fig.~\ref{fig4}d.
Such an enhancement is reminiscent of those observed in the AlAs quantum well
at much lower densities~\cite{de2000resistance} and can be related to the charge transport
along the domain wall loops near the first order Ising transition~\cite{jungwirth2001resistance}.
Possible evolution and hysteresis~\cite{de2000resistance,piazza1999first}
of such a peak in future experiments at lower temperatures may completely decipher the anticrossings.

Figure~\ref{fig4}e reveals the $R_{xx}$ behavior under tilted magnetic fields. Away from the anticrossing,
the $R_{xx}$ features near $\nu=$6 and 8 remain virtually unchanged for all accessible tilt angles.
This insensitivity arises from the spin splitting and pinning to the out-of-plane direction,
as explained above. By sharp contrast, on the two sides of the $\nu=7$ anticrossing,
the $R_{xx}$ peaks do evolve with the tilt angle.
The two peaks exchange their relative positions abruptly without elimilating the $\nu=7$ minimum  (dotted traces),
corresponding to the reversal of Ising polarization.
Such a tilt angle dependence implies that an in-plane magnetic field can couple the two LLs near their anticrossing,
modify the exchange interaction, and tune the valley polarization of quantum Hall Ising ferromagnet.

In conclusion, we fabricate high-quality layer-polarized {\it n}-type MoS$_2$ devices and, for the first time,
study valley-resolved SdH oscillations relevant to the spin-valley locked massive Dirac electron LLs.
With decreasing density, we observe LL crossings and valley ferrimagnet-to-ferromagnet transitions,
revealing a four fold enhancement of the valley Zeeman effect by the interactions.
Near integer ratios of the Zeeman-to-cyclotron energies,
we discover persistent LL anticrossings that can be attributed to the formation of quantum Hall Ising ferromagnets.
The Ising valley polarizations appear to be reversible by tuning the density and in-plane magnetic field.
Our results provide compelling evidence for strong many-body interaction effects
in the conduction bands of monolayer MoS$_2$ and establish a fertile ground for
exploring strongly correlated phenomena of massive Dirac electrons.

\begin{methods}

\noindent{\bf Sample preparation.} Bulk MoS$_2$ crystals were purchased from {2D Semiconductors}.
To form the hBN/MoS$_2$/hBN sandwich structure, we employed a well-developed dry transfer technique:
the MoS$_2$ flake was picked up by an hBN flake hold by a polymethyl methacrylate (PMMA) film,
which was then transferred onto a second hBN flake on a SiO$_2$/Si wafer~\cite{wang2013one,xu2016universal}.
The stacked heterostructure was annealed at 300~$^\circ$C
in an argon-protected environment to reduce trapped interfacial bubbles.
We used standard electron-beam (e-beam) lithography and reactive ion etching
to pattern the Hall-bar structure and selectively etch away the top hBN at the contact areas.
The e-beam lithography was performed again to define the electrodes.
Ti/Au (5~nm/60~nm) electrodes were coated with e-beam evaporation.
The number of layers of the MoS$_2$ flake was determined with optical contrast,
atomic force microscopy (AFM), and Raman spectroscopy (see Supplementary Information for details).

\noindent{\bf Electrical measurement.} A Stanford Research DS360 low-distortion function generator
was used to apply a 4.5~Hz, 2~mV source-drain bias voltage. The current was measured by
a Signal Recovery 7280 wide-bandwidth digital lock-in amplifier, and the longitudinal and Hall voltages
were measured by Stanford Research SR830 lock-in amplifiers with SR550 voltage preamplifiers.
The back gate voltage was provided by an Aim-TTi International PLH120 DC power supply.
Electrical measurements were performed using a 15~Tesla Oxford Instruments superconducting magnet system
with base temperature 1.5K, and using a 29~Tesla high-field system with base temperature 1.2~K at LNCMI-G.

\end{methods}

\bibliography{reference}

\begin{addendum}
\item This work is supported by the Research Grants Council of Hong Kong (Project No. 16300717, SBI17SC16),
HKUST-UoM Seed Fund 2017, and UT-Dallas Research Enhancement Fund.
We acknowledge the technical support from the Raith-HKUST Nanotechnology Laboratory
for the electron-beam lithography facility at MCPF and the support of the LNCMI-CNRS,
member of the European Magnetic Field Laboratory (EMFL).
\item[Author contributions]
N.W., F.Z., and J.L. conceived the project.
J.L. and T.H. fabricated and measured samples A,C, and sample B, respectively.
J.L. and B.A.P. conducted high-field experiments at LNCMI-G.
J.L. and F.Z. analyzed the data and the theoretical aspects.
J.L. performed the fittings and prepared the figures.
The remaining authors provided technical assistance in experimental measurements and analyses.
J.L., F.Z., and N.W. wrote the manuscript, which was proofread by all authors.
\item[Additional information]
Supplementary information is available in the online version of the paper.
Correspondence and requests for materials should be addressed to F.Z. or N.W.
\item[Competing Interests] The authors declare no competing financial interests.
\end{addendum}

\pagebreak
\noindent\textbf{\Large Figure 1}\vspace{0.45in}\\

\pagebreak
\noindent\textbf{\Large Figure 2}\vspace{0.45in}\\

\pagebreak
\noindent\textbf{\Large Figure 3}\vspace{0.45in}\\

\pagebreak
\noindent\textbf{\Large Figure 4}\vspace{0.45in}\\

\end{document}